\documentstyle[aps,epsf]{revtex}         

\begin{document}
%
%
%
\twocolumn[\hsize\textwidth\columnwidth\hsize\csname
@twocolumnfalse\endcsname

\title{Periodically-dressed Bose-Einstein condensates: a superfluid with an anisotropic and variable critical velocity}
\author{J. Higbie and D.M. Stamper-Kurn}
\address{Department of Physics, University of California, Berkeley CA  94720}
\date{\today }
\maketitle
\begin{abstract}
Two intersecting laser beams can produce a spatially-periodic
coupling between two components of an atomic gas and thereby
modify the dispersion relation of the gas according to a
dressed-state formalism. Properties of a Bose-Einstein condensate
of such a gas are strongly affected by this modification.  A
Bogoliubov transformation is presented which accounts for
interparticle interactions to obtain the quasiparticle excitation
spectrum in such a condensate.  The Landau critical velocity is
found to be anisotropic and can be widely tuned by varying
properties of the dressing laser beams.
\end{abstract}
\pacs{PACS numbers: ???}

]

In numerous physical systems, particles which are confined to a
medium can be treated as free particles whose properties are
modified by the medium.  For example, the dispersion relation of
electrons in a periodic solid is changed from the free-particle
relation to a band structure which is periodic in quasi-momentum
space.  Such electrons are then treated as particles with
properties which can be purposely engineered by modifying the
surrounding periodic structure.  In this Letter, we consider
similarly the task of engineering novel behaviour in a
quantum-degenerate atomic gas by placing the gas in a properly
constructed periodic medium.

A periodic potential for an atomic gas can be produced by
intersecting two or more laser beams.   A polarizable atomic gas
illuminated by two intersecting off-resonant laser beams  with
identical polarization and frequency but different wavevectors
experiences a spatially-periodic potential  proportional to the
light intensity, i.e.\ the atoms reside in a crystalline medium
made of light. Atoms in such media have been studied in the
non-degenerate \cite{madi97} and quantum degenerate
\cite{deng99talbot,burg01per,ande98atla,orze01squeeze} regimes. A
small frequency difference $\omega = \omega_1 - \omega_2$ can be
introduced between the laser beams to induce resonant Bragg
transitions between identical internal but different momentum
states.  Such Bragg transitions have been used to probe
Bose-Einstein condensates \cite{stam00leshouches}.

In this Letter, we present a different scheme for engineering
properties of an ultra-cold gas.  Rather than considering a
spatially periodic potential, we consider a spatially-periodic
coupling between two internal states of an atomic gas, i.e.\ we
consider a periodic medium contructed by laser beams which effect
Raman
 rather than Bragg transitions.  An atomic gas in this
medium is characterized by an easily tunable dressed-state
dispersion relation. In creating a Bose-Einstein condensate of
such a periodically-dressed atomic gas, one may explore how the
nature of free single-particle excitations can modify macroscopic
properties of a quantum fluid.  In particular, we develop a
Bogoliubov transformation which defines the quasi-particle
excitation spectrum, and show that this novel quantum fluid has a
tunable and anisotropic superfluid critical velocity.

\begin{figure}[htbf]
\epsfxsize=70mm \centerline{\epsfbox{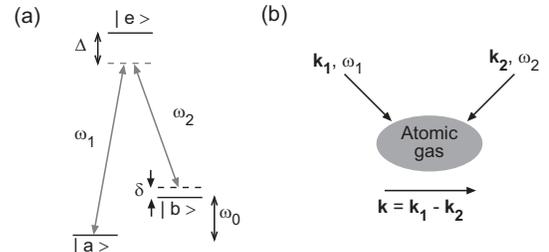}}
     \caption{Engineering propeties of a
     periodically-dressed atomic gas.  (a) Laser beams of
     frequency $\omega_1$ and $\omega_2$ may induce Raman
     transitions between internal states $|a\rangle$ and
     $|b\rangle$.  $\delta = (\omega_1 - \omega_2) - \omega_0$ is the detuning from the Raman
     resonance.  (b) Such a Raman transition imparts a momentum transfer of
     $\hbar {\bf{k}} = \hbar ({\bf{k}}_1 - {\bf{k}}_2)$, where ${\bf{k}}_1$ and
     ${\bf{k}}_2$ are the wavevectors of the Raman coupling
     lasers.
     }
     \label{fig:scheme}
\end{figure}

Let us consider a uniform dilute gas composed of atoms of mass $m$
with two internal ground states, $|a\rangle$ and $|b\rangle$,
separated by an energy $\hbar \omega_0$, and an excited state
$|e\rangle$ (see Fig.\ \ref{fig:scheme}).  This gas is exposed to
two laser beams (labeled 1 and 2), with wavevectors $\bf{k_1}$
and $\bf{k_2}$ and frequencies $\omega_1$ and $\omega_2$,
respectively.  The beams are polarized so that beam 1 connects
states $|a\rangle$ and $|e\rangle$ and beam 2 connects states
$|b\rangle$ and $|e\rangle$, while both beams have a large
detuning $\Delta$ from these allowed dipole transitions. These
laser beams can induce Raman transitions between the two internal
ground states.  A transition from state $a\rangle$ to $b\rangle$ :
results in a momentum kick of $\hbar {\bf{k}} = \hbar (
{\bf{k_1}} - {\bf{k_2}})$. Such coupling can be represented in
position space as a spatially-periodic coupling between the two
internal states, proportional to $e^{i {\bf{k}} \cdot {\bf{r}}}
|b\rangle \langle a |
 + e^{-i {\bf{k}} \cdot {\bf{r}}} |a \rangle \langle b |$.

Using second quantized notation, a gas which is continuously
illuminated by these beams can be described by the dressed-state
Hamiltonian \cite{cohe92}
\begin{eqnarray}
\mathcal{H} & = & \sum_{\bf{q}} \left( \left( \frac{\hbar^2 q^2}{2
m} -  \frac{\hbar \omega_0}{2}\right) a_{\bf{q}}^\dagger
a_{\bf{q}} + \left( \frac{\hbar^2 q^2}{2 m} + \frac{\hbar
\omega_0}{2}\right) b_{\bf{q}}^\dagger b_{\bf{q}} \right)  \nonumber \\
& + & \hbar \omega_1 \left( c_{1}^\dagger c_{1} - N_1 -
\frac{1}{2}\right) + \hbar \omega_2 \left( c_{2}^\dagger c_{2} -
N_2 - \frac{1}{2}\right) \nonumber \\
& + &  \sum_{\bf{q}} \left( \epsilon c_{2}^\dagger b_{\bf{q} +
\bf{k}/2}^\dagger c_{1} a_{\bf{q} - \bf{k}/2} + \epsilon^*
c_{1}^\dagger a_{\bf{q} - \bf{k}/2}^\dagger  c_2 b_{\bf{q} +
\bf{k}/2} \right) \label{eq:dressedhami}
\end{eqnarray}
Here $a_{\bf{q}}$ and $b_{\bf{q}}$ ($a_{\bf{q}}^\dagger$ and
$b_{\bf{q}}^\dagger$) are the annihilation (creation) operators
for atoms with wavevector ${\bf{q}}$  in internal states
$|a\rangle$ or $|b\rangle$, respectively. The operators $c_{1}$
and $c_{2}$ ($c_{1}^\dagger$ and $c_2^\dagger$), which are photon
annihilation (creation) operators for photons in beams 1 and 2,
can be replaced with $c$-numbers $\sqrt{N_1}$ and $\sqrt{N_2}$,
respectively, in the limit that the photon number is large.  We
then define $\Omega = 2 \epsilon \sqrt{(N_1 + 1)(N_2 + 1)} /
\hbar$ as the (real) two-photon Rabi frequency.

Now define single-atom states  $|a_{\bf{q}}\rangle \equiv |N_1 +
1; N_2; a, \bf{q} - \bf{k}/2 \rangle$ and $|b_{\bf{q}}\rangle
\equiv |N_1; N_2 + 1; b, \bf{q} + \bf{k}/2 \rangle$ which are
connected by a Raman transition.  The notation indicates that for
state $|a_{\bf{q}}\rangle$ there are $N_1+1$ photons in beam 1
and $N_2$ photons in beam 2, and the atom is in state $|a\rangle$
with momentum $\hbar {\bf{q}} + {\bf{k}}/2$, and similar for
state $|b_{\bf{q}}\rangle$.  These two states have the same total
momentum \cite{quasimomentumfootnote}, which we define as $\hbar
\bf{q}$ by a proper choice of reference frame.  The unperturbed
energies ($\epsilon \rightarrow 0$) of these states are
\begin{eqnarray}
\hbar \omega_{\bf{q}}^a &=& \frac{\hbar^2}{2 m} \left( \bf{q} -
\frac{k}{2}\right)^2 + \frac{\hbar \delta}{2} \\
\hbar \omega_{\bf{q}}^b &=& \frac{\hbar^2}{2 m} \left( \bf{q} +
\frac{k}{2}\right)^2 - \frac{\hbar \delta}{2}
\end{eqnarray}
where $\delta = (\omega_1 - \omega_2) - \omega_0$ is the detuning
from the Raman resonance.

Diagonalizing the Hamiltonian in the subspace of states
$|a_{\bf{q}}\rangle$ and $|b_{\bf{q}}\rangle$ yields the
dressed-state atomic eigenstates $|\pm_{\bf{q}}\rangle$ with
energies
\begin{equation}
\hbar \omega^\pm_{{\bf{q}}} = \frac{\hbar^2}{2 m} \left( q^2 +
\frac{k^2}{4} \right) \pm \frac{\hbar}{2} \sqrt{\left(\delta
-\frac{\hbar {\bf{q}} \cdot {\bf{k}}}{m}\right)^2 + \Omega^2}
\label{eq:dresseddispersion}
\end{equation}
and creation operators $\pi_{\bf{q}}$ (for state
$|+_{\bf{q}}\rangle$) and  $\mu_{\bf{q}}$ (for state
$|-_{\bf{q}}\rangle$) defined as
\begin{equation}
\left(\! \begin{array}{c} \pi_{\bf{q}} \\ \mu_{\bf{q}}
\end{array} \!\right) = \left(\! \begin{array}{c c} \cos
\theta_{\bf{q}}/2 & \sin \theta_{\bf{q}}/2 \\
- \sin \theta_{\bf{q}}/2 & \cos \theta_{\bf{q}}/2
\end{array} \!\right) \left( \! \begin{array}{c} a_{\bf{q}} \\ b_{\bf{q}} \end{array}
\! \right) \label{eq:dressedops}
\end{equation}
The dressed states are derived from the bare states
$|a_{\bf{q}}\rangle$ and $|b_{\bf{q}}\rangle$ by a rotation in
the two-dimensional subspace by the angle $- \theta_{\bf{q}}/2$
where $\tan \theta_{\bf{q}} = \Omega / (\delta + \hbar {\bf{q}}
\cdot {\bf{k}} / m)$.

\begin{figure}[htbf]
\epsfxsize=50mm \centerline{\epsfbox{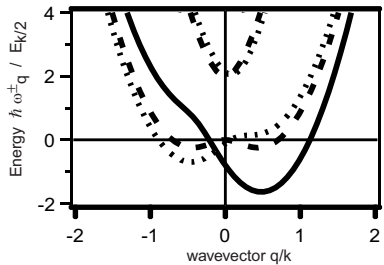}}
     \caption{Dressed-state dispersion relation: single particle energies $\hbar \omega^\pm_{\bf{q}}$
     vs.\ (normalized) wavevector $q/k$.  The excitation momentum $\hbar {\bf{q}}$ and
     the momentum transfer $\hbar {\bf{k}} = \hbar ({\bf{k_1}} - {\bf{k_2}})$
     are chosen to be colinear.  Energies are scaled by
     $E_{k/2} = \hbar^2 (k/2)^2/ 2 m$.
     For the traces shown, $\hbar \Omega / E_{k/2} = 2$, and $\hbar \delta / E_{k/2} = 3$ (solid lines), $\hbar \delta / E_{k/2} = 0$ (dashed lines),
     or $\hbar \delta / E_{k/2} = -1$ (dotted lines).
     For small $|\delta|$ (e.g.\ $\hbar \delta / E_{k/2} = -1$), the dispersion relation acquires a secondary minimum.
     As $\delta$ changes sign, the momentum of the lowest energy state changes discontinuously.
     For $ \delta = 0$, the motional ground state is
     degenerate.
     }
     \label{fig:dispfig}
\end{figure}

The dressed-state dispersion relation (Eq.\
\ref{eq:dresseddispersion}) is shown in Fig.\ \ref{fig:dispfig}.
The motional ground state occurs in the lower dressed state at a
wavevector $\bf{Q}$, which is near either $\pm {\bf{k}}/2$
depending on the sign of $\delta$.

Let us now consider the effects of Raman coupling on a
two-component Bose-Einstein condensate, which is now formed of a
macroscopic population of atoms in the lowest energy state
$|-_{\bf{Q}}\rangle$, i.e. the condensate is formed of atoms in a
coherent superposition of atomic states with differing momenta
$|a_{{\bf{Q}} - {\bf{k}}/2}\rangle$ and $|b_{{\bf{Q}} +
{\bf{k}}/2}\rangle$.  This condensate has a two-branch excitation
spectrum, reflecting the presence of two internal states (and,
correspondingly, two distinct dressed-state levels).  As with a
scalar Bose-Einstein condensate, weak repulsive interactions
should yield  phonon-like excitations at low energies and a
free-particle-like dispersion relation at high energies. However,
unlike for a scalar condensate, we expect properties of a
periodically-dressed Bose-Einstein condensate to reflect the
anisotropy of the dressed-state dispersion relation.

Let us now treat explicitly the effects of atomic collisions by
writing the many-body Hamiltonian as
\begin{equation}
{\mathcal{H}} = \sum_{\bf{q}} \left( \hbar \omega^-_{\bf{q}}
\mu_{\bf{q}}^\dagger \mu_{\bf{q}} + \hbar \omega^+_{\bf{q}}
\pi_{\bf{q}}^\dagger \pi_{\bf{q}} \right) +
{\mathcal{H}}_{\mbox{\small{int}}}
\end{equation}
Considering only elastic binary collisions (which conserve the
number of atoms in each of the internal states) characterized by
identical $s$-wave scattering lengths $a$, we may write the
interaction Hamiltonian ${\mathcal{H}}_{\mbox{\small{int}}}$ as
\begin{equation}
{\mathcal{H}}_{\mbox{\small{int}}} = \frac{g}{2} \sum_{\bf{q}}
\left(n_{\bf{q}} n_{-\bf{q}} - N\right)\label{eq:Hint}
\end{equation}
Here $g = (4 \pi \hbar^2 a/m) \times  V^{-1}$, $V$ is the volume
occupied by the gas, $N$ is the  number of atoms in the gas, and
$n_{\bf{q}}$, the spatial Fourier transform of the density
operator, is
\begin{eqnarray}
n_{\bf{q}} & = & \sum_{\bf{k}} \left( \begin{array}{c c}
a_{\bf{k} + \bf{q}}^\dagger & b_{\bf{k} + \bf{q}}^\dagger
\end{array} \right) \left(
\begin{array}{c} a_{\bf{k}} \\ b_{\bf{k}} \end{array} \right) \\
& = & \sum_{\bf{k}} \left( \begin{array}{c c}
\pi_{\bf{k} + \bf{q}}^\dagger & \mu_{\bf{k} + \bf{q}}^\dagger
\end{array} \right) {\mathcal{R}} \left( \frac{\theta_{\bf{k}} -
\theta_{\bf{k} - \bf{q}}}{2} \right) \left(
\begin{array}{c} \pi_{\bf{k}} \\ \mu_{\bf{k}} \end{array} \right)
\end{eqnarray}
Note that, since the rotation matrix ${\mathcal{R}}$ (defined in
accordance with Eq.\ \ref{eq:dressedops}) is not generally
diagonal, elastic collisions do not generally conserve the number
of atoms in the upper and the lower dressed states, respectively.

We now make use of the Bogoliubov approximation \cite{bogo47} in
which we assume a macroscopic population of $N_0$ atoms in the
lowest energy state of wavevector $\bf{Q}$, and substitute
$\mu_{\bf{Q}} = \mu_{\bf{Q}}^\dagger = \sqrt{N_0}$. To proceed,
we consider the 4-component vectors
\begin{equation}
v = \left( \begin{array}{c} \mu_{\bf{Q}+ \bf{q}} \\
i \mu_{\bf{Q} - \bf{q}}^\dagger \\
\pi_{\bf{Q}+\bf{q}} \\
i \pi_{\bf{Q} - \bf{q}} \end{array} \right), \, \mbox{ and } \, w
= \left( \begin{array}{c} \mu_{\bf{Q}+ \bf{q}}^\dagger \\ i
\mu_{\bf{Q} - \bf{q}} \\ \pi_{\bf{Q}+\bf{q}}^\dagger \\ i
\pi_{\bf{Q} - \bf{q}}^\dagger \end{array} \right)
\end{equation}
The Bose commutation relations of the dressed-state annihilation
and creation operators can be expressed as $[v_i, w_j] =
\delta_{ij}$.  Isolating terms of order $N^2$ and $N$, we may now
approximate the Hamiltonian as
\begin{equation}
{\mathcal{H}} \simeq \hbar \omega^-_{\bf{Q}} N + \frac{g}{2} (N^2
- N) + \sum_{\bf{q}\neq 0} \frac{w_i H_{ij} v_j}{2} + \hbar
\omega^+_{\bf{Q}} \pi^\dagger_{\bf{Q}} \pi_{\bf{Q}}
\end{equation}
where $H$ is a 4 $\times$ 4 matrix of the form $H_{ij} =
{\mathcal{E}}_{ij} + \mu x_i x_j$.  The diagonal matrix
${\mathcal{E}}_{ij}$ has entries $ {\mathcal{E}}_{\bf{q}}^-$,
$-{\mathcal{E}}_{-\bf{q}}^-$, ${\mathcal{E}}_{\bf{q}}^+$, and
$-{\mathcal{E}}_{-\bf{q}}^+$ where ${\mathcal{E}}_{\bf{q}}^\pm =
\hbar ( \omega^\pm_{{\bf{Q}}+{\bf{q}}} - \omega^-_{\bf{Q}})$.  The
chemical potential is given by $\mu = g N$, and we define
\begin{equation}
x = \left( \begin{array}{c} \cos \Delta_{\bf{q}} \\
-i \cos \Delta_{-\bf{q}} \\
\sin \Delta_{\bf{q}} \\
-i \sin \Delta_{-\bf{q}} \end{array} \right)
\end{equation}
where $\Delta_{\bf{q}} = (\theta_{\bf{Q} + \bf{q}} -
\theta_{\bf{Q}})/2$.

The quasi-particle energies and their creation and annihilation
operators are found by diagonalizing the matrix $H$.  Given the
invertible matrix $M$ for which $M H M^{-1} = \tilde{H}$ is
diagonal, allowing $v$ to transform as a column vector
($\tilde{v}_i = M_{ij} v_j$) and $w$ as a row vector
($\tilde{w}_j = w_i M^{-1}_{ij}$), we find the Bose commutation
relations $[\tilde{v}_i, \tilde{w}_j] = \delta_{ij}$ to be
preserved for the quasi-particle creation and annihilation
operators ($\tilde{\mu}_{\bf{q}}$, $\tilde{\mu}_{\bf{q}}^\dagger$,
$\tilde{\pi}_{\bf{q}}$, $\tilde{\pi}_{\bf{q}}^\dagger$) defined by
the elements of $\tilde{v}$ and $\tilde{w}$.  One finds the
diagonal elements of $\hbar^{-1} \tilde{H}$ to be $
\tilde{\omega}^-_{\bf{Q}+\bf{q}}$, $
-\tilde{\omega}^-_{\bf{Q}-\bf{q}}$,
$\tilde{\omega}^+_{\bf{Q}+\bf{q}}$, and
$-\tilde{\omega}^+_{\bf{Q}-\bf{q}}$ which define the lower and
upper quasi-particle excitation energies at wavevector ${\bf{Q}}
\pm \bf{q}$.

\begin{figure}[htbf]
\epsfxsize=50mm \centerline{\epsfbox{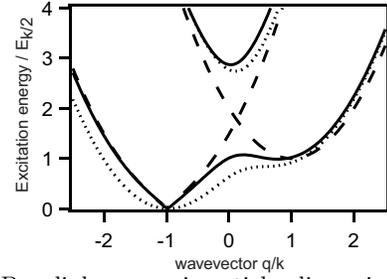}}
     \caption{Bogoliubov quasi-particle dispersion relation of a
     periodically-dressed Bose-Einstein condensate.  We consider a gas of $^{87}$Rb exposed to
     counter-propagating Raman beams near the
     $^{87}$Rb principal transition, with $\hbar \delta = - E_{k/2}$,
     $\hbar \Omega = 2 E_{k/2}$ and $\mu = 0.61 E_{k/2} = h \times$ 2.5 kHz.  Excitations parallel to $\bf{k}$
     are considered.  Lower and upper quasi-particle excitation branches
     are shown (solid lines) as a function of the wavevector $q$ of the excitation.  The quasi-particle
     energy spectrum is higher than the free dressed-state
     dispersion relation (dashed lines).  Dotted lines show the quasi-particle dispersion relation in the case of no Raman coupling
     ($\Omega \rightarrow 0$).  The effect of Raman coupling is to induce a level anti-crossing.
     }
     \label{fig:quasidisp}
\end{figure}

Figure \ref{fig:quasidisp} shows the quasiparticle spectrum
calculated for excitations parallel to the Raman transition
momentum transfer $\bf{k}$.  In choosing parameters for this
calculation, we have in mind an experimentally convenient
realization  with a Bose-Einstein condensate of $^{87}$Rb. One
may choose the internal ground hyperfine states $|a\rangle =
|F=1, m_F = -1\rangle$ and $|b\rangle = |F=2, m_F = 1\rangle$
which can be connected by a two-photon Raman transition. Choosing
these states has the benefit that the Raman transition frequency
is insensitive to magnetic field fluctuations, both hyperfine
states are magnetically trappable, and, specifically in
$^{87}$Rb, inelastic collisions are scarce. Furthermore, the
scattering lengths for all elastic collisions are nearly the
same, justifying the assumption made in Eq.\ \ref{eq:Hint}.  We
consider the case of counter-propagating Raman laser beams which
are nearly resonant with the $D_2$ transition ($k \simeq 4 \pi /
\lambda$ where $\lambda = 780$ nm). The chemical potential $\mu =
h \times$ 2.5 kHz corresponds to a condensate density of $3
\times 10^{14} \, \mbox{cm}^{-3}$.

As shown in the figure, the quasi-particle energies are higher
than the free dressed-state energies due to the repulsive
interactions between atoms.  Comparing the quasi-particle
spectrum to that  for the two-component condensate in the absence
of Raman coupling ($\Omega \rightarrow 0$), one sees that the
effect of the dressing lasers is to introduce avoided
level-crossings to the spectrum.  The spectrum for quasi-particle
excitations near the condensate momentum $\bf{Q}$ (i.e.\
wavevectors ${\bf{Q}} + {\bf{q}}$ for small $\bf{q}$) is linear,
describing phonon-like excitations. For excitations parallel to
the direction of momentum transfer and making small $q$
approximations to the Hamiltonian, we find the lower excitation
spectrum to have the limiting value $\hbar
\tilde{\omega}_{\bf{Q}+\bf{q}}^- = c^* \hbar q$ where $c^* =
\sqrt{\mu / m^*}$ is the Bogoliubov speed of sound corresponding
to an effective mass $m^*$ determined by the curvature of the
dressed-state dispersion relation at its minimum. We note further
that the complex matrix $H$ which appears in the Hamiltonian has
two positive and two negative real eigenvalues
\cite{stabilityfootnote}, as is required for the stability of the
condensate.

Finally, we consider the implications of the dressed-state
dispersion relation for the superfluidity of the
periodically-dressed Bose-Einstein condensate.  An explanation
for the dissipation-less flow of a superfluid below a critical
velocity $v_L$ was provided by Landau, who used kinematic
arguments to define $v_L = \min E({\bf{q}})/\hbar q$ where
$E(\bf{q})$ is the quasi-particle excitation energy at wavevector
$\bf{q}$ \cite{land41}. For weakly interacting scalar
Bose-Einstein condensates, the Bogoliubov quasi-particle
dispersion relation gives a Landau critical velocity $v_L = c$
which is equal to the speed of sound $c = \sqrt{\mu / m}$, and
the onset of scattering of microscopic impurities thus occurs by
the scattering of phonons.  This contrasts with the behaviour of
superfluid $^4$He in which excitations are phonon-like at low
momenta, while at higher momenta there exists a secondary minimum
in the excitation spectrum corresponding to rotons \cite{nozi90}.
The Landau critical velocity in that case is set by the roton
minimum \cite{meye61}.

\begin{figure}[htbf]
\epsfxsize=50mm \centerline{\epsfbox{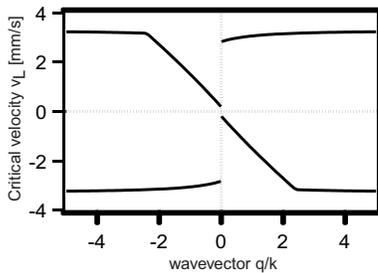}}
      \caption{Landau critical velocity in a periodically-dressed Bose-Einstein condensate.
     Velocities aligned with (positive $v_L$) or counter to (negative $v_L$) the Raman momentum transfer
     $\bf{k}$ are considered, and we take $\hbar \Omega = 2 E_{k/2}$.  At large detunings $|\delta|$,  $v_L$ has the same magnitude for flow in both
     directions, with a value approaching the Bogoliubov speed of sound $c = \sqrt{\mu/m} = 3.3$ mm/s for a
     $^{87}$Rb condensate at the density of $3 \times 10^{14} \, \mbox{cm}^{-3}$.  At smaller detunings, an anisotropy
     in $v_L$ develops.  For the regimes \{$\delta >0$, $v_L > 0$\} and \{$\delta<0$, $v_L < 0$\}, $v_L$ is determined by the speed of Bogoliubov sound modified
     by the effective mass $m^* > m$.  In the regimes \{$\delta>0$,$v_L<0$\} and \{$\delta<0$,$v_L>0$\},
     the magnitude of $v_L$ is lowered due to the secondary minimum in the dispersion relation.}
     \label{fig:landau}
\end{figure}

The presence of a secondary minimum in the dispersion relation of
periodically-dressed Bose-Einstein condensates suggests an analogy
to superfluid $^4$He and a reduction of the superfluid velocity
below the speed of sound.  Figure \ref{fig:landau} shows the
Landau critical velocity calculated for impurity velocities
parallel to the Raman momentum transfer $\bf{k}$, as the Raman
laser detuning $\delta$ is varied.  For large values of
$|\delta|$, the critical velocity is equal to the speed of sound
$\sqrt{\mu/m}$ as for a single-component Bose-Einstein
condensate.  As $|\delta|$ is lowered, the critical velocity
becomes anisotropic.  In one direction, $v_L$ is dramatically
lowered due to scattering at the secondary minimum (an
``artificial roton'') in the dispersion-relation.  We may
approximate the ``artificial roton'' minimum as occurring at
energy $\hbar |\delta|$ and quasi-particle momentum $\hbar k$,
obtaining $|v_L| \simeq |\delta / k|$ in this regime. Thus, the
superfluid properties can be controlled by varying the detuning
$\delta$ or by varying $k$ by changing the relative angle between
the Raman laser beams.  In the other direction, $v_L = c^* =
\sqrt{\mu/m^*}$ as determined by phonon scattering.

One may also vary the intensity of the Raman beams, thereby
changing $\Omega$.  An important impact of changing the Raman
coupling strength is not only altering the Landau critical
velocity (changing both the effective mass $m^*$ and the position
of the secondary minimum), but also controlling the degree to
which quasi-particles in the lower or upper excitation branches
can be created by scattering off an obstacle.  That is, the Landau
criterion determines the onset of dissipation for a moving
superfluid, but does not describe the \emph{strength} of such
dissipation. A large Raman coupling would enhance scattering into
secondary minimum of the lower excitation branch, while in the
limit $\Omega \rightarrow 0$, this scattering rate clearly
vanishes.  A detailed calculation of this dissipation rate
(similar to Refs.\ \cite{timm97symp,chik00,kett00coll}) will be
given elsewhere.

In summary, we have described a means of engineering a novel
quantum fluid  composed of dressed-state atoms in a
spatially-periodic Raman coupling medium.  This quantum fluid
should be amenable to study using current methods for probing
ultracold atomic gases.  The quasi-particle dispersion relation
can be probed by studying collective excitations (i.e.\ density
and magnetization modulations) at various length scales (see
reviews in \cite{generalvar}).   Aspects of superfluidity in
gaseous Bose-Einstein condensates have been investigated in
recent experiments \cite{madi99vort,chik00}, and similar methods
can be used to investigate the superfluidity of a
periodically-dressed condensate.  For example, the anisotropic
Landau critical velocity can be probed by studying the scattering
of an impurity gas which propagates through the fluid
\cite{chik00} .  Further theoretical work should address a number
of issues. For instance, the current treatment describes the
quantum fluid exclusively in momentum space. Experimentally, such
a fluid would be produced within a trapped volume.  Excitations
with wavelengths much smaller than the size of the fluid (as
probed with Bragg scattering \cite{stam00leshouches}) should be
well described by our treatment, while the description of longer
wavelength excitations will require further theoretical
considerations. Another interesting consideration is the
behaviour of the quantum fluid with the Raman lasers tuned
precisely to the Raman resonance, $\delta = 0$, at which point
the motional ground state is doubly degenerate.  We suspect that
in the presence of an external trapping potential, such a quantum
fluid would be described as evolving in a double-well potential
in momentum space.  Finally, our scheme may also be extended to
condensates with more than two internal states, such as spinor
Bose-Einstein condensates \cite{sten98spin}, giving greater
flexibility in engineering properties of gaseous quantum fluids.

J.H.\ acknowledges support from the NSF.

\bibliographystyle{prsty}

\begin{thebibliography}{10}

\bibitem{madi97}
K.W. Madison {\it et~al.}, App. Phys. B {\bf 65},  693  (1997);
M. Raizen, C. Salomon, and Q. Niu, Physics Today {\bf 50},  30
(1997); M.~Ben Dahan {\it et~al.}, Annales de Physique {\bf 23},
111 (1997).

\bibitem{deng99talbot}
L. Deng {\it et~al.}, Phys. Rev. Lett. {\bf 83},  5407  (1999).

\bibitem{burg01per}
S. Burger {\it et~al.}, Phys. Rev. Lett. {\bf 86},  4447  (2001).

\bibitem{ande98atla}
B.P. Anderson and M.A. Kasevich, Science {\bf 282},  1686  (1998).

\bibitem{orze01squeeze}
C. Orzel {\it et~al.}, Science {\bf 291},  2386  (2001).

\bibitem{stam00leshouches}
D.M.\ Stamper-Kurn, and W. Ketterle, arXiv:cond-mat/0005001.

\bibitem{cohe92}
C. Cohen-Tannoudji, J. Dupont-Roc, and G. Grynberg, {\em
Atom-Photon
  Interactions} (Wiley, New York, 1992).

\bibitem{quasimomentumfootnote}
If the laser beams are treated as an external pertubation rather
than as part
  of the quantum system, one should more properly speak of the quasi-momentum.

\bibitem{bogo47}
N.N. Bogoliubov, J. Phys. (USSR) {\bf 11},  23  (1947).

\bibitem{stabilityfootnote}
Taking the characteristic polynomial as $P(\lambda) = \det(H -
\lambda I)$, one
  finds (1) $P(\lambda \rightarrow \pm \infty) \rightarrow \infty$, (2) $P(0) >
  0$, and (3) by evaluating $P(\lambda)$ at plus or minus the dressed-state
  eigenenergies, one finds $P(\lambda)$ crosses zero for some $\lambda >0$ and
  $\lambda <0$. Thus, $P(\lambda)$ has four zeros on the real axis.

\bibitem{land41}
L.D. Landau, J. Phys. (USSR) {\bf 5},  71  (1941).

\bibitem{nozi90}
P. Nozi{\`e}res and D. Pines, {\em The Theory of Quantum
Liquids}, vol. 2 ed.
  (Addison-Wesley, Redwood City, CA, 1990).

\bibitem{meye61}
L. Meyer and F. Reif, Phys. Rev. {\bf 123},  727  (1961); D.R.
Allum, {\it et al.}, Phil. Trans. R.
  Soc. London A {\bf 284},  179  (1977).

\bibitem{timm97symp}
E. Timmermans and R. Cote, Phys. Rev. Lett. {\bf 80},  3419
(1997).

\bibitem{chik00}
A.C. Chikkatur {\it et~al.}, Phys. Rev. Lett. {\bf 85},  483
(2000).

\bibitem{kett00coll}
W. Ketterle, A.P. Chikkatur, and C. Raman, cond-mat/0010375.

\bibitem{generalvar}
{\em Bose-Einstein condensation in atomic gases}, {\em
Proceedings of the
  International School of Physics ``Enrico Fermi,'' Course CXL}, edited by M.
  Inguscio, S. Stringari, and C.E. Wieman (IOS Press, Amsterdam, 1999).

\bibitem{madi99vort}
K.W. Madison, {\it et al.}, Phys. Rev. Lett. {\bf
  84},  806  (2000);
M.R. Matthews {\it et~al.}, Phys. Rev. Lett. {\bf 83},  2498
(1999); C. Raman {\it et~al.}, Phys. Rev. Lett. {\bf 83},  2502
(1999).

\bibitem{sten98spin}
J. Stenger {\it et~al.}, Nature {\bf 396},  345  (1998).

\end{thebibliography}

\end{document}